# Isotope effect on the Casimir force

Lanyi Xie[1,4], Fuwei Yang[2,4] and Bai Song[1,3,4†]

[1]*Department of Energy and Resources Engineering, Peking University, Beijing 100871, China.*

[2]*Center for Nano and Micro Mechanics, Tsinghua University, Beijing 100084, China.*

[3]*Department of Advanced Manufacturing and Robotics, Peking University, Beijing 100871, China.*

[4]*Beijing Innovation Center for Engineering Science and Advanced Technology, Peking University, Beijing 100871, China.*

[†]Corresponding author. Email: songbai@pku.edu.cn

## ABSTRACT

Isotopic dependence of the Casimir force is key to probing new physics and pushing novel technologies at the micro and nanoscale, but is largely unexplored. In 2002, an isotope effect of $10^{-4}$ was estimated for metals—orders of magnitude beyond the experimental resolution. Here, by employing the Lifshitz theory, we reveal a significant isotope effect of over $10^{-1}$ for polar dielectrics. This effect arises from the isotope-mass-induced line shift of the zone-center optical phonons and is insensitive to the linewidth. We perform numerical analyses on both the imaginary and real-frequency axes, and derive analytical formulas for predicting the isotope effect.

Hendrik Casimir predicted in 1948 [1] that there exists an attraction between two perfectly conducting charge-neutral plates separated by a vacuum gap $d$. This so-called Casimir force per unit area is given by

$$F_C = -\frac{\pi^2 \hbar c}{240 d^4} = -1.3 \times 10^{-27} d^{-4} \text{ N m}^{-2}, \tag{1}$$

where $\hbar$ is the reduced Planck constant and $c$ is the speed of light. Similar to the van der Waals (vdW) force, the Casimir force also arises from quantum fluctuations of the electromagnetic (EM) field [2–8]. According to Casimir, the two plates form a cavity with high-reflectivity boundaries that limit the number of modes therein, leading to a net zero-point pressure that pushes inward [1]. The Casimir force finds broad applications in colloid and surface sciences, microsystems, and nanotechnologies due to the $d^{-4}$ dependence [9–11], and is of interest to several branches of physics including elementary particles, condensed matter, and cosmology [12–14].

Over the decades, much progress has been made in the measurement [15–22], computation [23–29], and understanding [30–33] of the Casimir force. In particular, a general theory for the Casimir interaction between objects of arbitrary materials, geometries, temperatures, and separating media was proposed by Lifshitz in 1956 based on the framework of fluctuational electrodynamics [23]. With the complex permittivities of the interacting bodies as the key input, this theory has since been widely adopted and laid the foundation for engineering both the strength and sign of the Casimir force. To this end, many schemes have been devised, such as using micro/nanostructures [34–37], metamaterials [38–40], phase transitions [41,42], topological insulators [42,43], tailored separating media [44–46], and external fields [46–48].

Among all the mechanisms for tuning the Casimir force, its isotopic dependence is rarely explored but of great interest, for example, in seeking the fifth force of nature [49–51]. Previous effort only estimated a negligible isotope effect of $10^{-4}$ for metals, which was two orders of magnitude below what was experimentally resolvable [49]. In contrast, large isotope effects on the vdW force has been extensively reported in diverse phenomena including vapor pressure, adhesion, and friction [52–55]. More importantly, an isotope effect of over $10^3$ on heat radiation from thermally fluctuating EM fields was recently revealed [56], which was attributed to the permittivity variation of polar dielectrics through isotope-mass-induced phonon line shift and broadening [57,58]. Together, these results motivate us to revisit the question as to how the Casimir force might change with the isotopic compositions of the interacting bodies.

In this work, we focus on the Casimir interactions between polar dielectrics [Fig. 1(a)] and highlight an isotope effect of over $10^{-1}$ for bulk plates, which is 3-orders-of-magnitude larger than that for metals. Reducing plate thickness generally enhances the isotope effect at large gaps but only by several folds. From the imaginary-frequency perspective, we attribute this significant isotopic dependence of the Casimir force to the atomic-mass-induced line shift of the zone-center phonons, and derive analytical formulas for predicting the isotope effect. Further, we provide additional insight on the real-frequency axis in terms of the surface phonon polaritons (SPhPs) modes. In contrast to thermal radiation [56], the phonon linewidth has a negligible influence. We conclude by showing that the difference between polar dielectrics and metals originates from the distinct isotopic-mass dependence of the phonon and plasma frequencies.

According to the Lifshitz theory [23], the Casimir force between two parallel plates at 0 K in vacuum can be expressed as

$$F(d) = \int_0^\infty F_\omega d\omega = -\frac{\hbar}{2\pi^2}\int_0^\infty d\omega \int_0^\infty kdk \ |\gamma_0(\omega,k)| \sum_{i=s,p} Z_i(\omega,k). \quad (2)$$

In practice, Eq. (2) holds well at finite temperatures since thermal effects only become appreciable at sufficiently large separations ($d \gg \hbar c/k_B T$, 7.6 μm at 300 K) [49]. Here, $F_\omega$ is the force spectrum, $k$ is the wavevector component parallel to the plates, $\gamma_0 = \sqrt{k_0^2 - k^2}$ is the perpendicular wavevector in vacuum with $k_0 = \omega/c$, and $Z$ is the exchange function given by

$$Z_{i=s,p}^{AB}(\omega,k) = \begin{cases} \text{Re}\left(\dfrac{-r_i^A r_i^B e^{2i\gamma_0 d}}{1 - r_i^A r_i^B e^{2i\gamma_0 d}}\right), & k \leq k_0, \\ \text{Im}\left(\dfrac{r_i^A r_i^B e^{2i\gamma_0 d}}{1 - r_i^A r_i^B e^{2i\gamma_0 d}}\right), & k > k_0, \end{cases} \quad (3)$$

where A and B denote the two plates, and $i$ refers to the $s$- and $p$-polarized modes. Both propagating and evanescent waves are included. For bulk plates, $r_i^A(\omega,k)$ and $r_i^B(\omega,k)$ are the Fresnel reflection coefficients at the plate-vacuum interfaces which are functions of the permittivity $\varepsilon(\omega)$ of the plates. Formulas for plates of finite thickness are provided in the Supplemental Material [59].

In light of the highly oscillating nature of the integrand in the real-frequency domain [23], $F(d)$ in Eq. (2) is usually expressed as a contour integration on the complex frequency plane along the imaginary-frequency axis $\omega = i\xi$:

$$F(d) = \int_0^\infty F_\xi d\xi = -\frac{\hbar}{2\pi^2}\int_0^\infty d\xi \int_0^\infty kdk \ |\gamma_0(i\xi,k)| \sum_{i=s,p} Z_i(i\xi,k). \quad (4)$$

Here, the permittivity $\varepsilon(i\xi)$ can be obtained via the Kramers-Kronig relation [29] as

$$\varepsilon(i\xi) = 1 + \frac{2}{\pi} \int_0^\infty \frac{\omega \text{Im}[\varepsilon(\omega)]}{\omega^2 + \xi^2} d\omega. \tag{5}$$

For polar dielectrics, the permittivity $\varepsilon(\omega)$ is given by the Lorentz model [56] as

$$\varepsilon(\omega) = \varepsilon_\infty \left(1 - \frac{\omega_{\text{LO}}^2 - \omega_{\text{TO}}^2}{\omega^2 - \omega_{\text{TO}}^2 + i\Gamma\omega}\right). \tag{6}$$

Here, $\varepsilon_\infty$ is the high-frequency permittivity and typically insensitive to isotopes; $\omega_{\text{TO}}$ and $\omega_{\text{LO}}$ are respectively the frequencies of the transverse and longitudinal zone-center optical phonons, which shift with the reduced mass $\mu$ of an isotope-engineered polar dielectric approximately as $\omega \propto \mu^{-1/2}$; and $\Gamma$ is the damping factor which increases with the isotopic mass fluctuations in ideal crystals but may vary considerably due to phonon-scattering by imperfections as the crystal quality varies [56].

To begin with, we study isotope-engineered cubic boron nitride (cBN) which is a prototypical polar dielectric and has already been used to study the isotope effects on heat conduction [60] and thermal radiation [56]. Usually, natural nitrogen is treated as isotopically pure (99.6% $^{14}$N), while the ratio between the two boron isotopes $^{10}$B and $^{11}$B is varied. Here, we consider cBN with four representative isotope ratios: c$^{11}$BN with pure $^{11}$B, c$^{\text{nat}}$BN with 20% $^{10}$B, c$^{\text{eq}}$BN with 50% $^{10}$B, and c$^{10}$BN with pure $^{10}$B. Their phonon properties can be found in our previous work [56]. Below we first focus on the isotope-induced phonon line shift and assume a fixed damping factor of 0.5 cm$^{-1}$.

In Fig. 1(b), we plot the calculated Casimir force between two c$^{11}$BN plates of representative thicknesses (bulk, 1 μm, 100 nm, and 10 nm) as a function of gap size. The same thickness is always assumed for both plates. We are concerned chiefly with the force magnitude so the negative sign is omitted unless otherwise noted. For bulk plates, $F$ increases from about 10 nN m$^{-2}$ at a 10 µm gap to 80 N m$^{-2}$ at a 10 nm gap. The scaling is roughly $d^{-4}$ at large gaps, since retardation is expected at distances comparable to the characteristic wavelengths of the plates [61]. At nanometer gaps, however, $F$ scales as $d^{-3}$. As the plates become thinner, the force decreases at large gaps but remains close to the bulk values for $d \ll t$. This is because the Casimir force between polar dielectrics at small gaps is expected to be dominated by the cavity SPhP modes [62,63], whose penetration depths decrease with and remain comparable to the gap size [64,65].

We further perform calculations for other isotopic configurations and notice that the Casimir force is always the weakest (denoted as $F_{\min}$) for the pair of c$^{11}$BN, which features the largest reduced mass and lowest phonon frequencies of any isotope-engineered cBN. (see Fig. S1 in [59]). With bulk c$^{11}$BN-c$^{11}$BN as the reference, we plot in Fig. 1(c) variations of the Casimir force $\Delta F = F - F_{\min}$ for the symmetric pairs of c$^{nat}$BN, c$^{eq}$BN, and c$^{10}$BN, and observe a monotonically increasing force as the fraction of $^{10}$B increases, regardless of the gap size. We also show the asymmetric case of c$^{11}$BN-c$^{10}$BN and find that the result overlaps well with that for c$^{eq}$BN-c$^{eq}$BN.

Based on the results above, we define the magnitude of the isotope effect for two plates of arbitrary isotopic compositions with respect to c$^{11}$BN-c$^{11}$BN as $\alpha = \Delta F/F_{\min} \times 100\%$. We focus on the isotope effect of c$^{10}$BN-c$^{10}$BN since it always yields the largest force. As shown in Fig. 1(d), $\alpha$ increases with decreasing gap size and saturates at a maximum value of about 2.8% for sufficiently small gaps regardless of the plate thickness, which is two-orders-of-magnitude larger than what was previously estimated for metals [49]. At larger separations, however, the isotope effect can be enhanced by several folds with thinner plates [59]. The largest enhancement appears at gap sizes around 4 μm.

To understand the isotope effect on the Casimir force, we perform spectral analysis on the imaginary-frequency axis in terms of the force spectra $F_\xi$, the cavity reflectivity $R^{AB} = r^A r^B$, and the permittivity $\varepsilon(i\xi)$. A small gap size of 10 nm is assumed for its larger isotope effect. As shown in Fig. 2(a), (b) and (c) respectively, all the $F_\xi$, $R^{AB}$, and $\varepsilon(i\xi)$ curves are very similar, with a flat region at low frequencies and a sharp drop at high frequencies where the cBN plates are effectively transparent [1,66]. In addition, the isotope-induced variations $\Delta F_\xi$, $\Delta R^{AB}$, and $\Delta\varepsilon(i\xi)$ with respect to c$^{11}$BN-c$^{11}$BN are also shown in Fig. 2 on the right axes. Notably, predominant peaks appear at around $2\times10^{14}$ rad/s, very close to $\omega_{TO}$ of c$^{11}$BN, indicating a crucial role of the zone-center phonons. These observations allow for a qualitative understanding of the gap-dependence of the isotope effect. First, we note that the Casimir force at gap size $d$ arises mainly due to waves of frequencies below $\xi_d = 2\pi c/d$. For large gaps with $\xi_d \ll \omega_{TO}$, the isotopes barely affect the force. As $d$ reduces, $\xi_d$ increases towards and passes $\omega_{TO}$, leading to a monotonically increasing $\alpha$. Eventually, $d$ becomes so small that $\xi_d \gg \omega_{TO}$ and no more isotopic enhancement is expected.

Mathematically, we can further establish analytical formulas for predicting the isotope effect at the small-damping limit. Details of the derivation can be found in [59]. Briefly, with $\Gamma \to 0$, $\text{Im}[\varepsilon(\omega)]/\omega$ approaches the Dirac $\delta$-function centered at $\sqrt{1+\beta}\omega_{TO}$, where $\beta$ denotes the isotope-induced relative phonon line shift [56] given by $\beta = \Delta\omega_{TO}/\omega_{TO} \approx \Delta\mu^{-0.5}/\mu^{-0.5}$. For $^{10}$B and $^{eq}$B, $\beta \approx$ 2.8% and 1.4%, respectively. With the Dirac $\delta$-approximation, the isotope effect on $\varepsilon(i\xi)$ can be written as

$$\varepsilon_\beta(i\xi) = \varepsilon\left(i\frac{\xi}{1+\beta}\right). \tag{7}$$

That is to say, a line shift $\beta$ in the real-frequency domain leads to a stretch of the imaginary-frequency permittivity by a factor of $1+\beta$. Based on Eq. (7), we can obtain an expression for the isotope-engineered Casimir force between identical plates:

$$F_\beta(d) = (1+\beta)^4 F[(1+\beta)d], \tag{8}$$

which agrees well with our exact calculations (Fig. S3). Recall that $F(d) \propto d^{-3}$ at small gaps, we then have $F_\beta(d) \approx (1+\beta)F(d)$ so the isotope effect is simply $\alpha \approx \beta$, as already demonstrated for c$^{10}$BN-c$^{10}$BN. As the gap size increases, the scaling law gradually approaches $F(d) \propto d^{-4}$ so $\alpha$ decreases and eventually goes to 0 with $F_\beta(d) \approx F(d)$. Eq. (8) also holds in the general case of pairing plates A and B, although the stretch factor is then $\sqrt{(1+\beta^A)(1+\beta^B)}$, which explains the similarity between c$^{11}$BN-c$^{10}$BN and c$^{eq}$BN-c$^{eq}$BN.

Despite the success on the imaginary-frequency axis, we have not been able to relate the isotope effect to the SPhPs of cBN, which are expected to dominate the Casimir force at small gaps [62,63]. To this end, we perform modal analysis in the real-frequency domain which is challenging in general but proves feasible in our case. The force spectra $F_\omega$ in Eq. (2) for various bulk pairs at $d$ = 10 nm are plotted in Fig. 3(a), which feature peaks and dips around the frequencies where $\text{Re}[\varepsilon(\omega)] = -1$, thus confirming the dominant role of the SPhPs in the near field. Here, the dips represent attractive forces (negative) due to the symmetric cavity SPhP modes as the SPhPs on the two vacuum interfaces couple, while the peaks indicate repulsion arising from the antisymmetric cavity modes [62]. By integrating $F_\omega$ across the reststrahlen band [Fig. 3(b)], we obtain net attractive forces in good agreement with those calculated using Eq. (4) for $d < 100$ nm (Fig. S4) [59]. In addition, we notice that the width of the highly reflective reststrahlen band scales as $(1 + \beta)(\omega_{LO} - \omega_{TO})$, which may offer a more intuitive picture for the magnitude of the isotope effect at small gaps.

With the fundamental role of the phonon line shift revealed, we proceed to a brief discussion of the line broadening. For perfect cBN crystals, the first-principles calculated damping factor $\Gamma$ increases with atomic mass fluctuations from ~0.5 cm$^{-1}$ for c$^{11}$BN and c$^{10}$BN to 1.3 cm$^{-1}$ for c$^{eq}$BN. However, $\Gamma$ may be much larger in real samples due to various imperfections [56]. In Fig. 4(a), we plot the Casimir force for pairs of bulk c$^{11}$BN, c$^{eq}$BN, and c$^{10}$BN at a 10 nm-gap versus $\Gamma$ from 0.5 to 100 cm$^{-1}$. All three force curves barely change at $\Gamma < 10$ cm$^{-1}$ (~0.01$\omega_{TO}$) and drop slightly for larger $\Gamma$ (~4% at 100 cm$^{-1}$). The corresponding isotope effect for c$^{10}$BN-c$^{10}$BN is also shown on the right axis, which remains almost constant. To explore the underlying mechanism, we plot in Fig. 4(b) the exchange functions for c$^{11}$BN-c$^{11}$BN and c$^{10}$BN-c$^{10}$BN at three representative $\Gamma$ (0.5, 1, and 2 cm$^{-1}$) with $k = 10^8$ m$^{-1}$ ($1/d$). Similar to $F_\omega$ in Fig. 3(a), $Z_p(\omega, k)$ features attractive dips and repulsive peaks, the widths of which broaden with $\Gamma$ while the heights reduce as $\Gamma^{-1}$. Therefore, the mode-resolved force remains the same upon frequency integration, which explains the negligible impact of $\Gamma$ on the Casimir force and its isotope effect. On the imaginary frequency axis, this insensitivity can also be anticipated by noting that Eq. (7) holds as long as $\Gamma$ is sufficiently small.

Apart from cBN, we also consider isotope-engineered lithium hydride (LiX with Li = $^6$Li, $^7$Li and X = H, D, T) which offers some of the largest phonon line shifts among polar dielectrics [56,67]. The isotope effects calculated via the Lifshitz theory for 5 symmetric LiX pairs and 15 asymmetric pairs at a 10-nm gap are shown in Fig. 5(a), with $^7$LiT-$^7$LiT as the reference and the average $\Delta\mu^{-0.5}/\mu^{-0.5}$ of the two plates on the x-axis. Indeed, the isotope effect for LiX is significant and reaches up to 55.6%. Further, we assume a hypothetical polar dielectric based on the permittivity model of LiX, and plot the exactly calculated isotope effect together with the prediction of Eq. (8) for small gaps ($\alpha \approx \beta \approx \Delta\mu^{-0.5}/\mu^{-0.5}$). All the data agree well with each other except for some minor deviations in a few asymmetric cases. In addition, we perform similar calculations for several real metals (Li, nickel and copper) and a hypothetical metal based on the Drude model of Li [56]. Negligibly small isotope effects on the order of $10^{-4}$ are obtained, in agreement with the estimation by Krause and Fischbach [59]. This is because for metals the plasma frequency shifts with the isotopic mass variation but with a reduction factor of about $10^{-3}$, that is, $\beta_\text{m} \approx \Delta m/m \times 10^{-3}$ [56].

In summary, we have systematically explored the isotope effect on the Casimir force between two plates in vacuum using the Lifshitz theory. In particular, we reveal a significant isotope effect of over $10^{-1}$ for polar dielectrics, which is three-orders-of-magnitude larger than the isotope effect for metals. This is understood via the reflectivity and permittivity variations along both the real and imaginary frequency axes, which originate from isotope-mass-induced zone-center phonon line shifts. Both the Casimir force and its isotope effect are insensitive to the phonon linewidth, in striking contrast to the case of thermal radiation [56]. Reducing the plate thickness enhances the isotope effect at large gaps but only by a few folds. The essential difference between polar dielectrics and metals lies in the distinct isotopic dependence of the phonon and plasma frequencies. We derive formulas that allow the isotope effect on the Casimir force to be accurately yet conveniently predicted from the relative frequency shift. Our work opens up new possibilities in engineering forces at small length scales and may also help probe new physics.

**ACKNOWLEDGMENT**

This work was supported by the National Natural Science Foundation of China (Grant No. 52076002), the Beijing Innovation Center for Engineering Science and Advanced Technology, the XPLORER PRIZE from the Tencent Foundation, and the High-performance Computing Platform of Peking University.

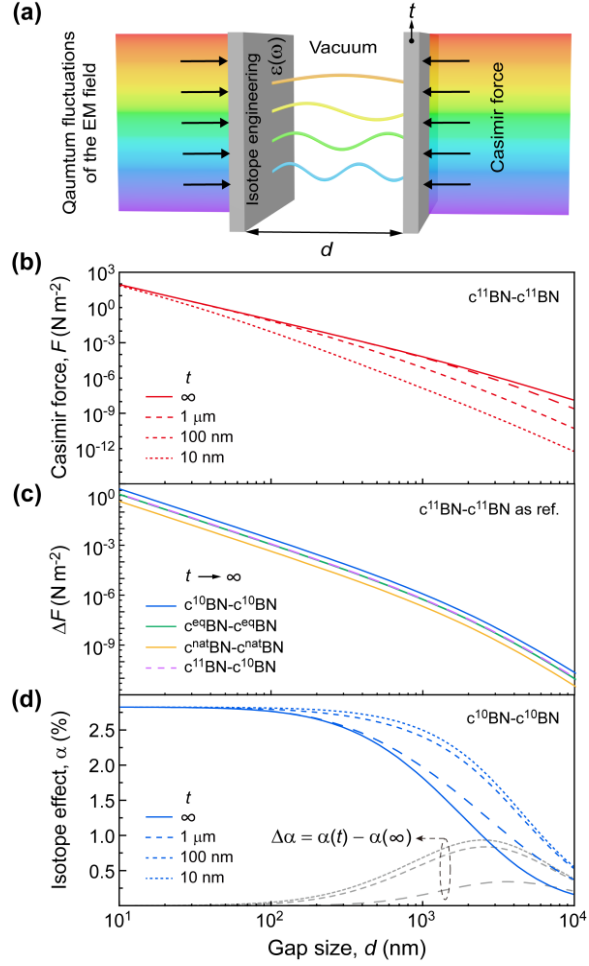

**FIG. 1. Isotope effect on the Casimir force between cBN plates.** (a) Schematic illustration. (b) Magnitude of the Casimir force for symmetric pairs of $c^{11}BN$ plates of varying thickness as a function of the gap size. (c) Force variation from bulk $c^{11}BN$-$c^{11}BN$ to symmetric pairs of $c^{nat}BN$, $c^{eq}BN$, $c^{10}BN$, and the asymmetric case of $c^{11}BN$-$c^{10}BN$. (d) Isotope effect for pairs of $c^{10}BN$ plates. Gray lines show the enhancement with reducing plate thickness.

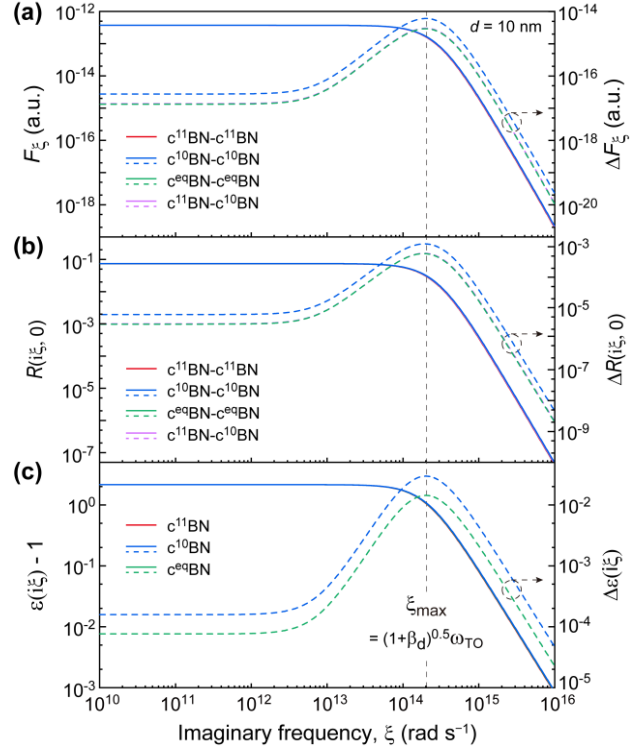

**FIG. 2. Understanding the isotope effect on the imaginary-frequency axis.** (a) Force spectra at a 10 nm-gap for pairs of $c^{11}BN$, $c^{eq}BN$, $c^{10}BN$, and $c^{11}BN$-$c^{10}BN$. (b) Corresponding reflectivity spectra at normal incidence. (c) Permittivities at imaginary frequency. Right axes show variations with respect to (pairs of) $c^{11}BN$ with the peak position marked.

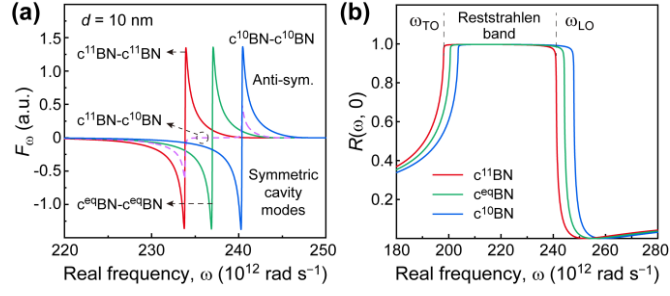

**FIG. 3. Understanding the isotope effect from the real-frequency perspective.** (a) Force spectra at a 10 nm-gap for pairs of $c^{11}BN$, $c^{eq}BN$, $c^{10}BN$, and $c^{11}BN$-$c^{10}BN$, showing the attraction (negative) and repulsion due to the symmetric and antisymmetric cavity SPhP modes, respectively. (b) Reflectivity spectra showing the isotope-induced shift and broadening of the Restrahlen band.

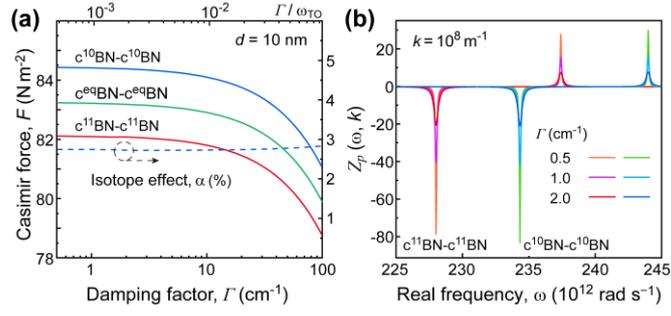

**FIG. 4. Influence of the damping factor.** (a) Casimir force for bulk $c^{11}BN$-$c^{11}BN$ and $c^{10}BN$-$c^{10}BN$ versus the damping factor $\Gamma$, and the corresponding isotope effect. (b) The exchange function at representative $\Gamma$ for $p$-polarized modes at $k = 10^8$ m$^{-1}$.

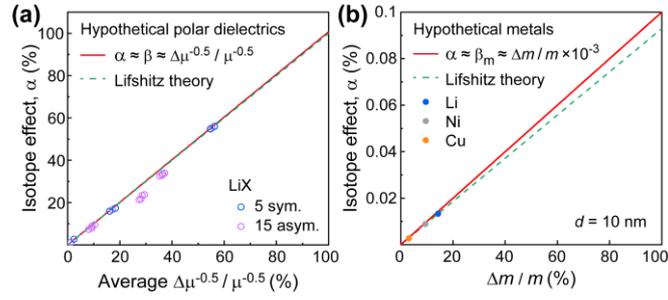

**FIG. 5. Predicting the isotope effect on Casimir force from atomic mass variations.** (a) Isotope effect for polar dielectrics as a function of $\Delta\mu^{-0.5}/\mu^{-0.5}$. Various combinations of LiX are plotted, together with a hypothetical material based on the Lorentz model of LiX. (b) Isotope effect for metals versus $\Delta m/m$. Real metals including lithium, nickel and copper are considered, together with a hypothetical metal based on the Drude model of Li. For both polar dielectrics and metals, the convenient estimations based respectively on the relative phonon and plasma frequency shift agree well with the exact calculations using the Lifshitz theory.